\documentclass{2023SCGE}

\usepackage{graphicx}	% Including figure files
\usepackage{amsmath}	% Advanced maths commands
\usepackage[table,xcdraw,dvipsnames]{xcolor}
\usepackage{hyperref}
\usepackage{float}
\usepackage{multirow}

\hypersetup{colorlinks=true,
           citecolor=cyan,
           linkcolor=cyan,
           filecolor=magenta,      
           urlcolor=cyan}
\DeclareGraphicsExtensions{.pdf,.png,.jpg}

\DeclareRobustCommand{\ion}[2]{%
\relax\ifmmode
\ifx\testbx\f@series
{\mathbf{#1\,\mathsc{#2}}}\else
{\mathrm{#1\,\mathsc{#2}}}\fi
\else\textup{#1\,{\mdseries\textsc{#2}}}%
\fi}

\usepackage[normalem]{ulem}

\newcommand{\czrm}[1]{{\bgroup\markoverwith{\textcolor{blue}{\rule[0.5ex]{2pt}{1pt}}}\ULon{#1}}}
\begin{document}
\ensubject{subject}
\ArticleType{Article}
\SpecialTopic{SPECIAL TOPIC: }
\Year{2023}
\Month{January}
\Vol{66}
\No{1}
\DOI{??}
\ArtNo{000000}
\ReceiveDate{January xx, xxxx}
\AcceptDate{April xx, xxxx}

\title{From Large Telescopes to the MUltiplexed Survey Telescope (MUST)}

\author[1]{Zheng Cai \thanks{Email: \href{mailto:zcai@tsinghua.edu.cn}{zcai@tsinghua.edu.cn}}}{}%
\author[1]{\ \ Song Huang\thanks{Email: \href{mailto:czhao@tsinghua.edu.cn}{shuang@tsinghua.edu.cn}}}{}%
\author[1]{\ \ Yu Liu\thanks{Email: \href{mailto:liuyu9@tsinghua.edu.cn}{liuyu9@tsinghua.edu.cn}}}{}%
\author[1]{\ \ Cheng Zhao\thanks{Email: \href{mailto:czhao@tsinghua.edu.cn}{czhao@tsinghua.edu.cn}}}{}%
\author[2]{\ \ Lei Huang }{}%

\AuthorMark{Cai, et al.}

\AuthorCitation{Cai, et al.}

\address[1]{Department of Astronomy, Tsinghua University, Beijing 100084, China}
\address[2]{Department of Precision Instrument, Tsinghua University, Beijing 100084, China}

\abstract{
Recent advances in astronomical observations have ushered in an era of remarkable discoveries. We now probe the Universe through multi-messenger signals, image the sky with unprecedented depth and resolution, and investigate individual sources using powerful large-aperture telescopes. Yet, a critical gap persists: the lack of wide-field, highly multiplexed spectroscopic capabilities needed to fully exploit the wealth of imaging data from current and upcoming surveys. In this review, we trace the historical development of large optical telescopes and spectroscopic surveys, assess the capabilities of ongoing and near-future facilities, and motivate the need for next-generation Stage-V spectroscopic experiments. As a representative example, we present the MUltiplexed Survey Telescope (MUST)—the first Stage-V spectroscopic facility currently under construction. MUST is a 6.5-meter telescope designed to obtain optical spectra for over 20,000 targets simultaneously within a $\sim$5 deg$^2$ field, using a modular focal plane populated with 6.2-mm pitch fiber-positioning robots. Over an 8-year survey in the 2030s, MUST aims to build the most comprehensive 3D spectroscopic map of the Universe to date, measuring redshifts for over 100 million galaxies and quasars and opening new windows into cosmology, Galactic structure, and time-domain astrophysics.
}

\vspace{1cm}

\vspace{0.2cm}

\maketitle

\tableofcontents

\begin{multicols}{2}

\section{Introduction}
    \label{sec:intro}

    Among all the branches of natural sciences, astronomy and astrophysics cover the largest scale of space and time. Historically, the invention of the telescope over 400 years ago transformed humanity's view of the world and the Universe, and triggered the first industrial revolution. Over time, with the aid of more advanced instrumentation, astronomers have uncovered numerous discoveries that have fundamentally altered humanity's understanding of the Universe. From the discovery of the cosmic microwave background radiation (CMB) to the establishment of the Big Bang cosmology, from the first detection of a planet outside of our Solar System to the characterization of the whole exoplanet family in the `statistical mode'\footnote{\url{https://pressbooks.cuny.edu/astrobiology/chapter/exoplanets-discoveries/}} \cite{2024arXiv240409143K}, all these impressive achievements testify to the ingenuity and ambitions of not only the astronomers, but also the engineers and technicians behind all the meticulously designed instruments that enable these discoveries. In the new millennium, the increasingly intimate collaboration between science and technology has pushed astronomy into more exciting frontiers. In 2015, 
    on the 100th anniversary of Einstein's discovery of his general relativity theory, gravitational waves were detected for the first time \cite{2016PhRvL.116f1102A}. In 2017, the first gravitational wave signal from a binary neutron star merger was also detected, accompanied by electromagnetic wave signals spanning almost the entire electromagnetic band \cite{2017PhRvL.119p1101A, 2017ApJ...848L..12A, 2017ApJ...848L..13A, 2017ApJ...848L..16S}. In 2018, the IceCube neutrino observatory at the South Pole captured a neutrino signal from a $\gamma{}$-ray blazer - the high-energy jet from a distant supermassive black hole \cite{2018Sci...361.1378I}. Now, astronomy has officially entered the `multi-messenger era', where humans can finally observe the Universe using all the carriers of fundamental information, e.g., electromagnetic waves, cosmic rays, gravitational waves, and neutrinos. 

    In this exciting new era, optical and near-infrared observations of the universe continue to be the cornerstone of astrophysics and cosmology. Over the past few decades, the astronomical community has pushed the design and development of large observing facilities to increasingly challenging limits on the ground or in space [e.g., James Webb Space Telescope (JWST) \cite{2006SSRv..123..485G}, Hubble Space Telescope (HST)\footnote{\url{https://www.stsci.edu/hst}}, Very Large Telescope (VLT)\footnote{\url{https://www.eso.org/public/teles-instr/paranal-observatory/vlt/}}, Keck Observatory twin telescopes (Keck)\footnote{\url{https://keckobservatory.org/}}, etc.]. Right now, several 30-40 meter ground-based telescopes, such as the Extremely Large Telescope (ELT) \cite{2023ConPh..64...47P}, Thirty Meter Telescope (TMT) \cite{2013JApA...34...81S, 2015RAA....15.1945S}, Giant Magellan Telescope (GMT) \cite{2012SPIE.8444E..1HJ}, are either under construction or being planned. With the help of revolutionary active and adaptive optic technologies, these new facilities provide astronomers with unprecedented high angular resolution and sensitivity capability across a broad range of wavelengths. All these mentioned facilities are often referred to as the ``general-purpose'' telescopes, which are usually equipped with many specialized instruments that cover a wide range of scientific capabilities and can carry out observations using different and flexible strategies. They excel at enhancing the details of carefully selected targets within a limited field of view (FoV; typically under a few tens of arcminutes), ranging from a remote quasar to an exoplanet.  

    Meanwhile, the development of astrophysics and cosmology also demands a different category of telescope —one that is optimized to efficiently collect information for a large number of targets over a wide field of view. This ``survey mode'' observation has a long history in astronomy. And, starting in the 21st century, more and more ground-based and space telescopes are being customized to enhance their survey capabilities for various scientific goals. For example, the Large Sky Area Multi-Object Fiber Spectroscopic Telescope (LAMOST; \cite{2012RAA....12..723Z}) has surveyed 10 million Milky Way stars. William Herschel Telescope Enhanced Area Velocity Explorer (WEAVE), a wide-field, massively multiplexed spectroscopic survey facility for the William Herschel Telescope users to obtain up to 1000 spectra of astronomical objects in a single exposure \cite{2024MNRAS.530.2688J}. Today, survey telescopes have become an indispensable component of modern astronomy. They represent astronomy in the golden age of information and "big data" and become engines of discovery for the community. They help us map the Universe, provide comprehensive statistical depictions of all types of astronomical targets, and can efficiently reveal the ``needle in the haystack'' - rare objects or anomalies that often possess valuable information. 

    More importantly, modern survey telescopes are critical in addressing many of the most fundamental questions in astronomy and physics. What is the nature of dark matter and dark energy? How did the inflationary process take place? How did the first generation of stars, galaxies, and galaxy clusters form? What is the detailed process of the structure formation? Are there habitable planets and life beyond our Solar System? The answers to these questions, the ``unknown unknowns'' they help uncover during the process, will lead to a new revolution in science. And, during this endeavor, survey telescopes will be at the center of many ambitious scientific projects. In particular for cosmology, extensive surveys of galaxies \cite{2004ApJ...606..702T, 2005MNRAS.362..505C} and supernovae \cite{1999ApJ...517..565P, 1998AJ....116.1009R, 1999Sci...284.1481B} in the large-scale structures of the Universe confirm that the baryonic matter represents only a tiny fraction of the overall matter density and the Universe is accelerating in expansion under the impetus of dark energy. Suffice it to say that survey telescopes have established themselves as a foundational pillar of modern observational cosmology.

    In recent years, the astronomy community has benefited from the data, scientific discoveries, and legacies of many successful surveys. For wide-field imaging surveys, the Sloan Digital Sky Survey (SDSS; \cite{2000AJ....120.1579Y, 2006AJ....131.2332G, 2016AJ....151...44D}), Dark Energy Survey (DES) \cite{2016MNRAS.460.1270D}, Hyper Suprime-Cam (HSC) Subaru Strategic Program (SSP) \cite{2018PASJ...70S...4A}, and the Kilo-Degree Survey (KiDS; \cite{2013ExA....35...25D, 2013Msngr.154...44D}) not only delivered high-quality multi-band images of (almost) the whole sky, but also enabled the supernova, galaxy clusters, and weak gravitational lensing aspects of the Stage-II and Stage-III experiments designed by the Dark Energy Task Force (DETF) report \cite{2006astro.ph..9591A}. Very soon, the start of the Euclid \cite{2011arXiv1110.3193L, 2024arXiv240513491E} satellite and the Rubin Observatory’s Legacy Survey of Space and Time (LSST) \cite{2018arXiv180901669T, 2019ApJ...873..111I, 2022ApJS..258....1B}, along with the launch of the China Space Station Telescope (CSST) \cite{2011SSPMA..41.1441Z, 2019ApJ...883..203G} and the Nancy Grace Roman Space Telescope (Roman) \cite{2015arXiv150303757S} will revolutionize our view of a deep and dynamical Universe. LSST alone is expected to catalog over 20 billion galaxies, while Euclid and CSST will map billions more, providing ``Stage-IV'' photometric redshifts and weak lensing data to advance cosmology \cite{2019ApJ...883..203G, 2019ApJ...873..111I, 2024arXiv240513491E}.
    
    However, the progress of these exciting imaging surveys also forces astronomers to face another critical problem: an outstanding lack of spectroscopic survey capability in the 2030s to 2040s. In astrophysics, spectroscopy measures the radial velocity, chemical abundances, dynamical properties, and many other physical properties of different targets. For cosmology, spectroscopy provides the precise redshift information to map the large-scale structure of the Universe. Starting from the early 2000s, a series of modern spectroscopic surveys, from the spectroscopic component of the SDSS \cite{2000AJ....120.1579Y, 2006AJ....131.2332G, 2016AJ....151...44D}, LAMOST \cite{2012RAA....12..723Z}, to Dark Energy Spectroscopic Instrument (DESI) \cite{2016arXiv161100036D}, Hobby-Eberly Telescope Dark Energy Experiment (HETDEX) \cite{2021ApJ...923..217G}, Subaru Prime Focus Spectrograph (PFS) \cite{2014PASJ...66R...1T}, 4-metre Multi-Object Spectroscopic Telescope (4MOST) \cite{2019Msngr.175....3D}, and Spectro-Photometer for the History of the Universe, Epoch of Reionization, and Ices Explorer (SPHEREX) \cite{2014arXiv1412.4872D}, have defined the current landscape of spectroscopic surveys. Among them, the DESI \cite{2016arXiv161100036D, 2016arXiv161100037D, 2019AJ....157..168D, 2022AJ....164..207D}, PFS, and 4MOST will measure the redshifts of $\sim 30$--40 million galaxies and observe a similar number of stars in the next decade. Together, they represent the spectroscopic requirements for Stage-IV cosmology. However, there is a dramatic three orders of magnitude gap between the available spectroscopic data in the 2030s and the number of records in the photometric catalogs \cite{2016arXiv161100036D, 2012SPIE.8446E..0TD}. This stark disparity underscores a pressing need for wide-field spectroscopic facilities to unlock the full potential of the survey data and the answers to more demanding cosmological questions.

    Motivated by this, multiple next-generation wide-field spectroscopic telescopes [e.g., MUST \cite{Zhang2023ConceptualDO, 2024arXiv241107970Z, 2025PASP..137b5001Z}, Spec-S5 \cite{2025arXiv250307923B}, Wide-field Spectroscopic Telescope (WST) \cite{2019BAAS...51g..45E, 2024arXiv240305398M}, and Eastern Spectroscopic Survey Telescope (ESST) \cite{2024SCPMA..6779511S}, etc.] have been proposed in recent years. Among these, MUST -- a 6.5-meter telescope with a 3$^\circ$ FoV and 20,000 robotic fiber positioners -- is the first one under construction. It plans to survey 100 million galaxies and quasars in the 2030s. Similarly, Spec-S5, planned for a 2037 first light, will carry out a full-sky redshift survey using two 6-meter telescopes. With a survey efficiency 10 times greater than DESI \cite{2024arXiv241107970Z, 2025arXiv250307923B}, MUST and Spec-S5 will extend our pursuit of dark matter, dark energy, nature of gravity, neutrino mass, and other fundamental questions about primordial cosmology into ``Stage-V''. And, eventually, ESST or WST could lead us into the era of spectroscopic surveys using 10-12 meter telescopes with a field of view (FoV) of about 3 square degrees,  and the %unprecedented 
    multiplexed capabilities could be close to 50,000, and occupied with different instruments, enabling transformative discoveries in cosmology, galaxy evolution, stellar population, time-domain astrophysics, and many more fields. For instance, WST will measure radial velocities and chemical abundances for millions of stars to advance our understanding of Milky Way formation; meanwhile, its collaboration with LSST and Euclid will enhance time-domain studies in astrophysics \cite{2024arXiv240305398M}. And, altogether, these 6-12-meter spectroscopic survey facilities can finally bridge the image-spectroscopy gap in the 2040s. 
    
    In this work, we provide a brief review of spectroscopic survey telescopes, with a focus on the ongoing MUST project. In \S\ref{sec:large-telescope}, we introduce the development of modern large telescopes and spectroscopic survey telescopes. In \S\ref{sec:must}, we use MUST as an example to discuss the properties and science goals of new-generation spectroscopic telescopes. Finally, the concluding remarks are made in \S\ref{sec:conclusion}.

\begin{table*}[ht]
\centering
\caption{The global large telescopes with primary mirror apertures larger than 6\,m.}
\label{tab:large_telescopes}
\resizebox{0.83\linewidth}{!}{
\begin{tabular}{@{}lcllc@{}}
\toprule
\textbf{Telescope}     & \textbf{Aperture} & \textbf{Function}         & \textbf{Site}                    & \textbf{Built}           \\ \midrule
E-ELT                 & 39 m             & General                   & Chile                           & under construction        \\
TMT                   & 30 m             & General                   & Hawaii, USA                     & under construction        \\
GMT                   & 25 m             & General                   & Chile                           & under construction        \\
LSST                  & 8 m              & Imaging Survey            & Chile                           & under construction        \\
JWST                  & 6.5 m            & General                   & NASA                            & launch                   \\
Keck I \& II          & $2 \times 10$ m  & General                   & Mauna Kea, Hawaii, USA          & 1994/1997                \\
GTC                   & 10.4 m           & General                   & Island La Palma, Spain          & 2008                     \\
HET                   & 9.1 m            & General                   & Mt Fowlkes, USA                 & 1998                     \\
SALT                  & 9.1 m            & General                   & Cape Town, South Africa         & 2006                     \\
LBT I \& II           & $2 \times 8.4$ m & General + interference    & Mount Graham, Arizona, USA      & 2005                     \\
Subaru                & 8.2 m            & Survey                    & Mauna Kea, Hawaii               & 2001                     \\
TAO                   & 6.5 m            & General                   & Chile                           & under construction        \\
VLT (4 $\times$ 8 m)  & $4 \times 8$ m   & General + interference    & Cerro Paranal, Chile            & 2002                     \\
Gemini I \& II        & $2 \times 8$ m   & General                   & Hawaii/Cerro Pachon, Chile      & 2002                     \\
Magellan I \& II      & $2 \times 6.5$ m & General + interference    & Las Campanas, Chile             & 2000/2003                \\
MMT                   & 6.5 m            & General                   & Mt Hopkins, USA                 & 2001                     \\
LZT                   & 6 m              & General                   & British Columbia, Canada        & 2004                     \\ \bottomrule
\end{tabular}}
\footnotesize
\\* \textbf{Full names:} 
Extremely Large Telescope (E-ELT), 
Thirty Meter Telescope (TMT), 
Giant Magellan Telescope (GMT), 
Legacy Survey of Space and Time (LSST), 
James Webb Space Telescope (JWST), 
W. M. Keck Observatory (Keck I \& II), 
Gran Telescopio CANARIAS (GTC), 
Hobby-Eberly Telescope (HET), 
Southern African Large Telescope (SALT), 
Large Binocular Telescope (LBT), 
Subaru Telescope, 
Tokyo Atacama Observatory (TAO), 
Very Large Telescope (VLT), 
Gemini Observatory, 
Magellan Telescopes, 
Multiple Mirror Telescope (MMT), 
Large Zenith Telescope (LZT).
\end{table*}

\section{Development of Modern Large Telescopes and the Spectroscopic Telescopes}
    \label{sec:large-telescope}

\subsection{Modern Large Telescopes}

    In the era of the 1970s, a few 4\,m telescopes, equipped with optical and infrared images, spectrographs, fueled the explosive growth of observational astronomy \cite{2008eiad.book.....M}. By the late 1970s, scientists turned their attention to segmented mirrors and larger monolithic mirrors to enhance light-gathering capabilities. During this period, Jerry Nelson played a key role in the development of splicing mirrors \cite{Nelson1985TheDO, Nelson2013}. A breakthrough in segmented mirrors was achieved on the Keck 10\,m telescope constructed in 1995. This telescope incorporates 36 hexagonal sub-mirrors. Meanwhile, a group of scientists led by Roger Angel have focused on new techniques for creating large monolithic mirrors, and now the largest monolithic mirror ranges from 6–8 meters\footnote{\url{https://www.lbto.org/}} \cite{1996SPIE.2807..354A}. 

    Meanwhile, the instrumentation of large telescopes is also rapidly developing. In the development of spectrographs, small prism instruments were replaced by larger grating instruments at the Cassegrain focuses to meet the need for higher spectral resolution. At the largest telescopes (e.g., TMT and VLT), most of the larger instruments are located at the Nasmyth focus on a platform that rotates with the telescope. Almost all spectroscopic instruments and imaging cameras now use solid-state electron detectors with high quantum efficiency, which, in combination with these telescopes, allow long-term focusing and observation of extremely faint objects \cite{Cheng2003}. 
    In 2025, nearly 30 telescopes with primary mirror apertures larger than 6\,m worldwide (cf.\,Table\,\ref{tab:large_telescopes}). Most of them are general-purpose telescopes.  

\begin{table*}[ht]
\centering
\caption{
    Partial ground-based and spatial optics Infrared telescopes [from the Table 6 of \cite{2010AN....331..338T} (redacted)].
    \textbf{Note:} $^1$ Name of the telescope; $^2$ Number of papers published in 2008 under the support of the telescope platform; $^3$ Number of citations of 2008 papers in 2009; $^4$ Average number of citations per paper in 2008. Although the SDSS (\textcolor{black}{blue box}) telescope has only a 2.5 m aperture, it has gained the highest scientific impact beyond imagination due to its large field of view multi-fiber spectral survey capability.
}
\label{tab:round-based-telescopes}
\resizebox{0.3\linewidth}{!}{
\begin{tabular}{@{}lccc@{}}
\toprule
\textbf{Telescope$^1$} & \textbf{Papers$^2$} & \textbf{Citat.$^3$} & \textbf{C/P$^4$} \\ \midrule
HST      & 206.6 & 765  & 3.70 \\
VLT      & 139.1 & 452  & 3.25 \\
Keck     & 59.6  & 333  & 5.59 \\
CFHT     & 38.0  & 152  & 4.00 \\
Gemini   & 34.3  & 108  & 3.15 \\
Subaru   & 33.0  & 138  & 4.18 \\
AAT      & 23.0  & 83   & 3.61 \\
Magellan & 16.4  & 68   & 4.15 \\
MMT      & 14.7  & 63   & 4.29 \\
WIYN     & 11.6  & 42   & 3.62 \\
TNG      & 9.3   & 15   & 1.61 \\
INT      & 8.2   & 43   & 5.24 \\ \midrule
\rowcolor{blue!20} 
SDSS     & 133.0 & 863  & 6.49 \\
2MASS    & 136.2 & 479  & 3.52 \\ \bottomrule
\end{tabular}}
\end{table*}

\subsection{The Current Status of Survey Telescopes}
    \label{sec:survey-telescope}

    The utilization of imaging techniques and spectroscopic observations emerged as the two most pivotal observational tools. In the forthcoming decade, the field is expected to witness significant advancements with the advent of cutting-edge survey telescopes that will significantly surpass the current depth of observations. The astronomy community has already employed large optical/infrared telescopes of the 4-8\,m class to implement image surveys.

    In particular, the 4.0\,m telescope Blanco (Dark Energy Survey, DES) \cite{2015AJ....150..150F, 2016MNRAS.460.1270D} and the 8.2\,m telescope Subaru (Hyper Suprime-Cam Imaging Survey, HSC) \cite{2004PASJ...56..381I, 2018PASJ...70S...1M, 2018PASJ...70S...4A} have made significant contribution to deep imaging survey in optical from $\sim 0.4-1\ \mu$m. Additionally, projects such as the 8.4\,m LSST\footnote{\url{https://rubinobservatory.org/}}, 2.4\,m Nancy Roman Telescope\footnote{\url{https://science.nasa.gov/mission/roman-space-telescope/}}, and 2.0\,m CSST\footnote{\url{https://www.nao.cas.cn/csst/}} are also in the construction phase. Overall, by 2030, these facilities could provide more than 30 billion galaxies with images. 

    While imaging surveys are relatively unchallenging in terms of the equipment and time required, spectroscopic survey speed is much lower, and such survey machines are much more demanding. In terms of scientific research, multiple large-field-of-view, multi-object spectroscopy survey machines are required to provide a more complete understanding of the nature of new objects. For example, only through spectroscopic surveys can one obtain detailed physical parameters of the objects, including redshift, radial velocity, stellar chemical composition, and galaxy kinematics, among others. The evolution of spectroscopic surveys has been a cornerstone in advancing our understanding of stars, galaxies, and the universe's large-scale structure. From the pioneering efforts of the SDSS \cite{2000AJ....120.1579Y, 2006AJ....131.2332G, 2016AJ....151...44D} through its successive phases, to modern initiatives like DESI \cite{2016arXiv161100036D}\footnote{\url{https://www.desi.lbl.gov/}} and WEAVE \cite{2024MNRAS.530.2688J}\footnote{\url{https://weave-project.atlassian.net/wiki/spaces/WEAVE/overview}}, and looking ahead to future projects, these surveys have systematically gathered invaluable spectral data that underpin significant discoveries in astronomy and cosmology.

\subsection{SDSS Wide-field Spectroscopic Surveys}
    \label{sec:spectroscopic_surveys}

    SDSS, launched in 2000 as its first phase, aimed to create the most detailed three-dimensional maps of the Universe. Its 2.5-meter telescope has a 3$^\circ$ FoV and is equipped with 640–1,000 optical fibers \cite{2006AJ....131.2332G}. It focused on imaging and spectroscopy of approximately one million objects, including galaxies, quasars, and stars. From 2000 to 2008, the SDSS provided a comprehensive catalog that has been fundamental for studies of galactic structure and the large-scale distribution of matter \cite{2000AJ....120.1579Y}. From 2008--2014, SDSS-III ushered a new era in the spectroscopic survey, the Baryon Oscillation Spectroscopic Survey (BOSS) aimed to measure baryon acoustic oscillations (BAO) with high precision to constrain cosmological models, significantly improving our understanding of dark energy and the expansion history of the Universe \cite{2013AJ....145...10D}. From 2014 to 2020, SDSS-IV further expanded the survey’s capabilities with projects such as the extended BOSS (eBOSS) and the Mapping Nearby Galaxies at Apache Point Observatory (MaNGA). The eBOSS aimed to refine cosmological parameters by mapping the distribution of galaxies and quasars over a larger volume, enhancing the precision of BAO measurements \cite{2016AJ....151...44D, 2021PhRvD.103h3533A}. MaNGA focused on integral field spectroscopy of around 10,000 nearby galaxies, allowing detailed studies of their internal structures, star formation processes, and evolutionary states \cite{2015ApJ...798....7B}. 

    SDSS represents a triumph for spectroscopic telescopes. As evidenced by \cite{2010AN....331..338T}, SDSS, despite its modest aperture of 2.5\,m, has surpassed both the Hubble Space Telescope and most substantial ground-based optical infrared telescopes in terms of both paper number and paper citation (cf.\,Table\,\ref{tab:round-based-telescopes}). This is attributable to its capacity to conduct rapid spectroscopic surveys covering 7 square degrees and utilizing thousands of optical fibers. It is also noteworthy that the SDSS was constructed and initiated at the beginning of this century, and its instruments have been continuously upgraded to implement the most advanced survey projects for two decades. For instance, SDSS has completed four phases of survey projects, including SEGUE-1/2, BOSS, eBOSS, APOGEE-1/2, and MaNGA. As indicated in the ESO planning report \cite{2010AN....331..338T}, the significant success of SDSS has imparted two key insights:

    \begin{enumerate}
        \item Large-field-of-view, multi-object spectroscopic survey telescopes 
        can maintain competitiveness and substantial productivity over the course of decades.
        \item The advent of data volume with more than an order of magnitude bigger than before can engender unanticipated scientific surprises.
    \end{enumerate}

\begin{table*}[ht]
\centering
\caption{
    Comparison of performance parameters of existing and future spectroscopic survey telescopes. %(some of the data in the table are from Table A3 in the literature \cite{2016Msngr.165...39E}). 
    \textbf{Note:} Some data are updated from the projects' websites.}
\label{tab:performance}
\resizebox{\textwidth}{!}{
\begin{tabular}{|>{\raggedright\arraybackslash}p{6cm}|>{\centering\arraybackslash}p{2.5cm}|>{\centering\arraybackslash}p{2.5cm}|>{\centering\arraybackslash}p{2.5cm}|>{\centering\arraybackslash}p{3cm}|>{\centering\arraybackslash}p{2.5cm}|>{\centering\arraybackslash}p{2.5cm}|}
\hline
\rowcolor{gray!20}
\textbf{Telescope and Survey Project Name} & \textbf{Starting Time} & \textbf{Main Mirror Area $A$/m$^2$} & \textbf{Field of View $\Omega$/deg$^2$} & \textbf{Number of Fibers} & \textbf{Proportion of Time Spent on Patrols $f_{\text{use}}$} & \textbf{Patrol Performance Indicators $\alpha/\alpha_{\text{DESI}}$} \\ \hline
SDSS-I/II            & 2000 & 4.91  & 7.07  & 640   & 0.5  & 0.034 \\ \hline
SDSS-III             & 2008 & 4.91  & 7.07  & 1 000 & 0.5  & 0.053 \\ \hline
VISTA/4MOST          & 2025      & 13.20 & 4.00  & 2 436 & 0.7  & 0.76  \\ \hline
Mayall/DESI          & 2020      & 12.57 & 7.50  & 5 000 & 0.5  & 1.00  \\ \hline
VLT/MOONS            & 2025      & 52.81 & 0.14  & 1 000 & 0.3  & 0.016 \\ \hline
Subaru/PFS           & 2025      & 52.81 & 1.10  & 2 400 & 0.25 & 0.30  \\ \hline
\rowcolor{red!20}
MUST                 & 2029–     & 33.18 & 6.20  & 20 000 & 0.7  & 20.5  \\ \hline
\end{tabular}
}
\end{table*}

\subsection{New-generation Wide-field Spectroscopic Surveys}
    \label{sec:next-gen}

\begin{figure*}
\centering 
\includegraphics[width=0.95\textwidth]{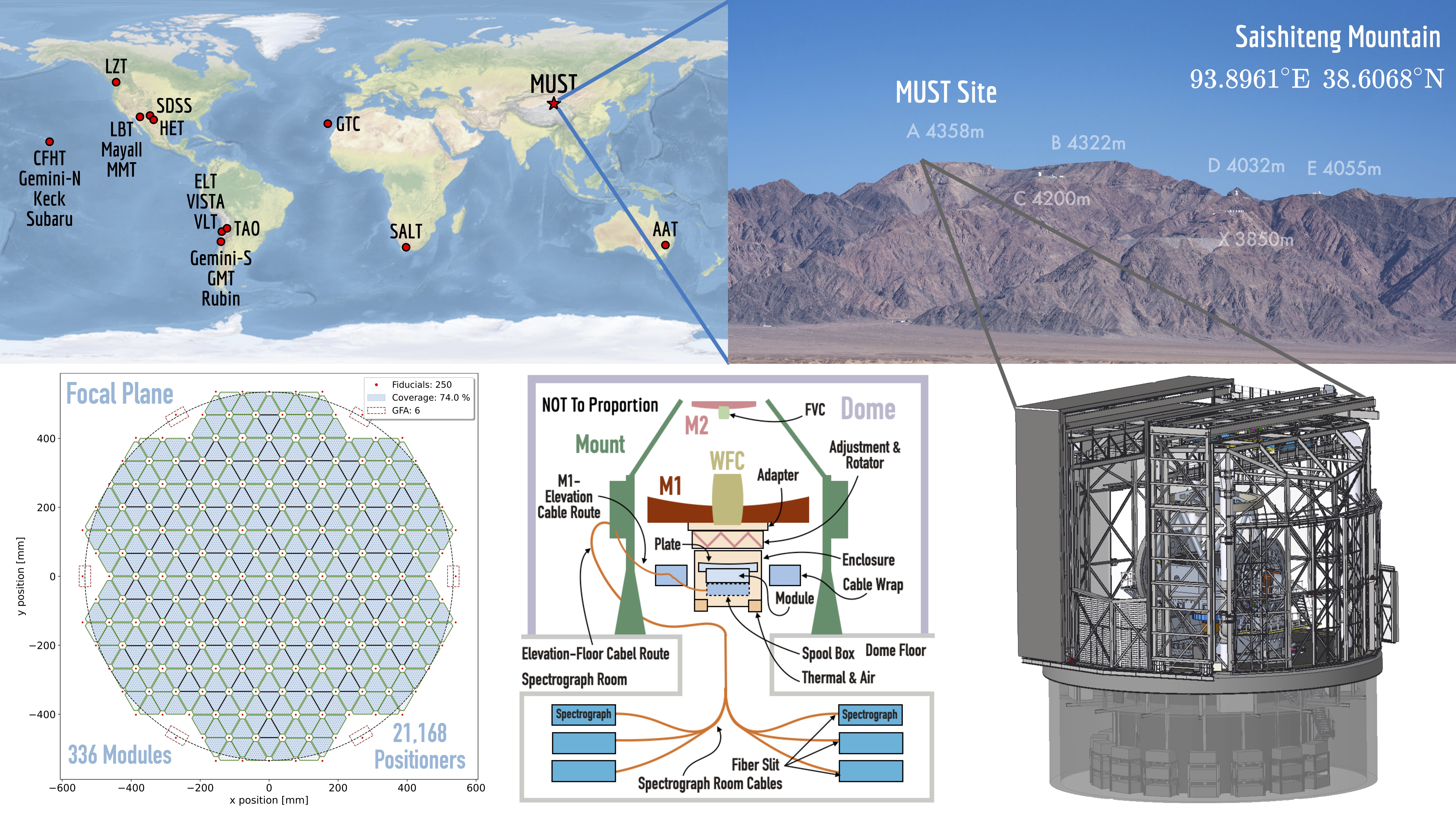}
\caption{
    Overview of the MUST project. Top left: Location of the MUST site in Qinghai province of China, shown in comparison with other major telescope sites worldwide. Top right: A picture of Saishiteng mountain near Lenghu town in Haixi Mongol and Tibetan autonomous prefecture, Qinghai province. The highest peak – Peak A, with an altitude of 4,358 m - was selected as the site for MUST. From right to left, the bottom panels illustrate the schematic of conceptual design of telescope overview, optical system (also cf.\,Figure\,\ref{fig:Schematic_Optical_design}(b)), %modular focal plane, robotic fiber positioner, 
    and spectrograph system layout (also cf.\,Figure\,\ref{fig:spectrograph}), and the fiducial design of the modular focal plane of MUST. According to the current design, MUST will host $>$20,000 robotic fiber positioners utilizing 336 triangular modules \cite{2024arXiv241107970Z}.}
\label{fig:must_overview}
\end{figure*}
    
    From 2020--2025, at least five telescopes are incorporated into spectroscopic surveys, including three 4\,m class telescopes, e.g., WHT-4.2m/WEAVE, Mayall-4m/DESI, 4-MOST, and two 8\,m telescopes, including Subaru-8m/PFS in Japan, VLT-8m/MOONS in Europe \cite{2014SPIE.9147E..0NC, 2020Msngr.180...10C}. 

    WEAVE, installed on the William Herschel Telescope (WHT), represents a next-generation spectroscopic survey instrument designed to address a broad spectrum of astronomical questions \cite{2024MNRAS.530.2688J}. Capable of simultaneously observing up to 1,000 objects, WEAVE is optimized for both wide-field and high-resolution spectroscopy. Its primary goals include studying stellar populations and dynamics within the Milky Way, exploring the formation and evolution of galaxies, and investigating the distribution of dark matter \cite{2012SPIE.8446E..0PD}. WEAVE’s versatility enables detailed examinations of individual stars and large-scale surveys of galaxies, bridging the gap between stellar astronomy and cosmology. Looking ahead, several ambitious spectroscopic surveys are poised to further our comprehension of the cosmos. 

    Building on the legacy of SDSS, DESI obtained spectra in 2021 \cite{2022AJ....164..207D}. DESI has embarked on an ambitious five-year survey to explore the nature of dark energy with spectroscopy of 40 million galaxies and quasars, aiming to determine redshifts and employ BAO at redshifts up to 3.5 \cite{2013arXiv1308.0847L, 2022AJ....164..207D}. Also, the growth of structure is expected to be measured, and the modifications to general relativity are to be tested. DESI has a 3-deg diameter wide-field corrector that focuses the light onto 5020 fiber positioners on the 0.8\,m diameter focal surface. Roughly 500 fibers are fed into a petal, and each petal is connected to a spectrograph via a 50\,m fiber cable bundle. The ten spectrographs split the light into three channels that overall cover the wavelength of 360-980\,nm with a resolution of 2000 to 5000. Till now, DESI has constructed the largest and most precise 3D map of the universe. DESI's first-year data have provided unprecedented precision, with measurements of the Universe's expansion history, achieving a current accuracy of nearly or slightly better than 1\% \cite{2024arXiv240403000D, 2024arXiv240403002D, 2025JCAP...01..124A}.

    The Prime Focus Spectrograph (PFS) on the Subaru Telescope is currently conducting its commissioning \cite{2024SPIE13096E..05T}. PFS is a powerful multi-object spectrograph designed to take advantage of Subaru’s 1.3 deg$^2$ field of view at its prime focus \cite{2014PASJ...66R...1T, 2016SPIE.9908E..1MT}. Developed by an international collaboration led by the Kavli Institute for the Physics and Mathematics of the Universe (Kavli IPMU) and the National Astronomical Observatory of Japan (NAOJ), PFS is engineered to collect spectra of up to 2,400 astronomical objects simultaneously. This capability further increases the efficiency of data collection compared to single-object spectrographs. Its spectroscopic coverage extends from the optical into the near-infrared (roughly 380\,nm to 1.3\,$\mu$m), allowing the study of key emission and absorption features that trace everything from stellar populations to the chemical enrichment of galaxies across cosmic time \cite{2016SPIE.9908E..1MT}. Scientifically, PFS is crucial for tackling some of the most pressing questions in astrophysics and cosmology. Studies of stellar dynamics and elemental abundances in the Milky Way and nearby galaxies provide insight into the processes of galaxy formation and evolution. Technically, to accomplish these goals, PFS uses a system of robotic fiber positioners that rapidly and precisely place fibers onto targets in the focal plane, feeding light to high-throughput spectrographs located off the telescope. This combination of a massive multiplex advantage (thousands of fibers) and broad wavelength coverage positions PFS as a key instrument for new-generation cosmological surveys and detailed studies of galaxy assembly \cite{2014PASJ...66R...1T}. 
    
    In the future, DESI-II is a planned extension of the Dark Energy Spectroscopic Instrument (DESI) survey, aiming to operate from 2029 to 2035. DESI-II will push DESI’s cosmological measurements into new domains by increasing galaxy density and redshift coverage ($z\lesssim3$) while serving as a stepping stone toward a Stage-V facility (Spec‐S5) \cite{2022arXiv220903585S}. It will utilize hardware upgrades, such as the proposed new skipper‐CCDs with readout noise lower than 1 electron, improved mirror cooling, and motor-controlled collimators in the spectrographs, which can help extend the survey to fainter galaxy populations and improve the survey speed by 50\%. Such improvements can help DESI-II expand into the $z > 2$ regime by observing Lyman-break galaxies (LBGs) and Ly$\alpha$ emitters (LAEs) at $2<z<3.5$. DESI-II will also emphasize its synergy with the LSST project. While the planned $\sim$5,000 deg$^2$ footprint and the 3 million redshifts at $z > 2$ fall short of the requirements of a Stage-V experiment, they will help us learn valuable lessons and better prepare for a true Stage-V survey. Shortly after DESI-II's debut, MUST will help us officially complete the transition from Stage-IV to Stage-V, thanks to its significantly higher survey efficiency at $z > 2$, which is made possible by MUST's vastly improved light-collecting and multiplex capabilities. By measuring up to 40 million redshifts at $2<z<5.5$, MUST will help realize the cosmological potential embedded in the high-redshift large-scale structure.

    Table\,\ref{tab:performance} provides a comprehensive overview of the key parameters for both existing and forthcoming major spectroscopic surveys, along with their performance metrics. While the combined efforts of previous and ongoing spectroscopic surveys—along with newly commissioned instruments—are expected to yield close to 50 million spectra of extragalactic sources by 2030, the availability of medium-to-high resolution and medium-to-high signal-to-noise spectra still lags far behind the rapid growth rate of galaxy images. Nevertheless, the so-called ``stage-V" spectroscopic survey demands a performance that is more than 10 times that of the DESI survey \cite{2022arXiv220903585S, 2022arXiv221109978C, 2024arXiv241107970Z}, which currently stands as the most substantial survey to date. Therefore, a Phase-V spectroscopic survey is in high demand in astronomy. In this context, MUST \cite{Zhang2023ConceptualDO, 2024arXiv241107970Z, 2025PASP..137b5001Z} is a representative example of the ``Stage-V" survey. 

\section{Multiplexed Survey Telescope (MUST)}
    \label{sec:must}

\subsection{General Goals of MUST}
    \label{ssec:general-goal}

\begin{figure*}
\centering 
\includegraphics[width=0.95\textwidth]{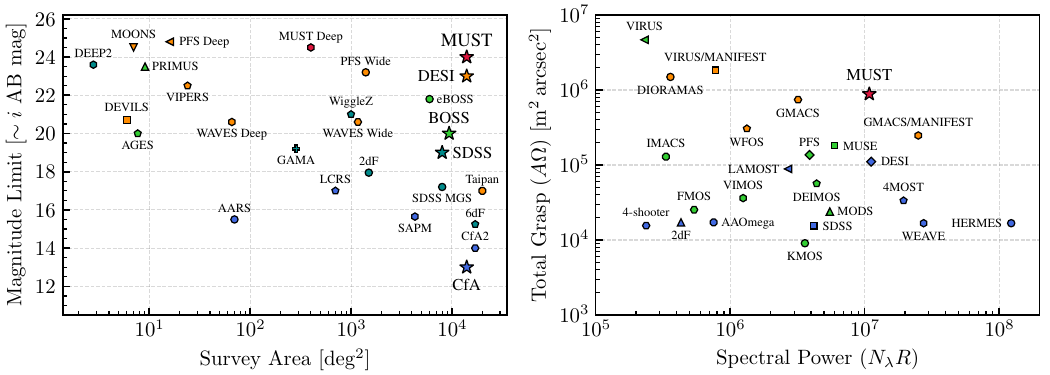}
\caption{
    Comparison of spectroscopic surveys and instruments on large telescopes. Left: Survey sky area and apparent magnitude limit in $i$-and reachable by different redshift surveys. Major galaxy spectroscopic surveys representing different stages of cosmological observations are highlighted as stars. MUST strives to become the first Stage-V survey in operation by the early 2030s. Right: Total Grasp vs. Spectral Power for different spectroscopic instruments (see \cite{2014AdOT....3..265H}). The total grasp is defined as the product of area and solid angle ($A\Omega$) for photon collection, while spectral power is the product of spectral resolution and the number of spectral resolution elements. MUST is designed to occupy a region of parameter space optimal for both cosmology and time-domain science. 
}
\label{fig:redshift_magnitude}
\end{figure*}

    Multiple reports have suggested that the most significant and versatile scientific instrument to complement the next-generation wide-area imaging surveys is the multi-object spectroscopic survey telescope \cite{2016arXiv161001661N}. As a wide-field multi-object spectroscopic survey telescope led by Tsinghua University, MUST will have an aperture of 6.5\,m and a field of view of 6.2 square degrees, thus addressing a significant gap in the current telescope parameter space. This telescope is currently planned to be situated on the highest peak (i.e., Peak A) on Saishiteng mountain near Lenghu town in the Haixi Mongol and Tibetan autonomous prefecture, Qinghai province (cf.\,Figure\,\ref{fig:must_overview}). With a fiber count over 20,000, a spectral resolution of more than 3,000, and spectral coverage ranging from 0.3 to 1.0\,$\upmu$m, MUST may be the first Stage-V spectroscopic survey telescope, with a spectral acquisition capability of more than 10 times that of the current Stage-IV facilities (e.g., DESI \cite{2016arXiv161100036D, 2022AJ....164..207D}, etc.) (cf.\,Figure\,\ref{fig:redshift_magnitude}).

    Large-field-of-view multi-object spectroscopic surveys have been a powerful tool for cosmological studies since the 1990s \cite{1998AJ....116.3040G, 2000AJ....120.1579Y, 2016arXiv161100036D}. The spatial distribution of large samples of galaxies, quasars, and the intergalactic medium, as obtained by surveys, can directly depict the three-dimensional large-scale structure of the Universe and its evolution over time. This provides abundant experimental data for testing cosmological models, measuring the densities of dark matter and energy, and revealing the rate of expansion of the Universe and the geometry of space, among other things. The BOSS and eBOSS surveys of SDSS have already comprehensively mapped the cosmic space at $z<1$ \cite{2015ApJS..219...12A, 2016AJ....151...44D, 2017MNRAS.470.2617A, 2021MNRAS.500..736B}. The ongoing and forthcoming Stage-IV dark energy surveys (e.g., DESI, PFS) are dedicated to the cosmological study of the Universe at $0.7<z< 2.2$. The early Universe at $z>2$ will be left to next-generation surveys like MUST. 

    A key scientific goal of MUST is to improve the cosmology of the Universe at $0.7<z<2.2$ by conducting large-scale redshift surveys at $2<z<5$, a first-time exploration of the expansion history of the early Universe, and by investigating the physical mechanisms that drove the initial `bloom' and subsequent `expansion' of the Universe \cite{2024arXiv241107970Z}. MUST will initiate a series of large-scale spectroscopic surveys within the initial five years of its establishment. These surveys are poised to catalyze groundbreaking advancements in several domains of astrophysics, including gravitational wave cosmology, time-domain astronomy, the detection of primordial inflation and the nature of dark energy, the formation of first-generation galaxies and clusters of galaxies, the detection of exoplanets and evidence of habitable Earths, and the history of structure and formation of the Milky Way, etc.

\subsection{Examples of MUST Key Science}
    \label{ssec:key-sci-example}

    As a next-generation spectroscopic survey telescope, MUST will significantly enhance our capacity to map the 3-D Universe. By collecting hundreds of millions of spectra from galaxies, stars, and transient sources in the 2030s, MUST is expected to enable significant scientific advances — from precision cosmology to galactic archaeology — and to address fundamental questions in cosmology, high-energy physics, galaxy evolution, and time-domain astrophysics. This subsection outlines several key scientific objectives enabled by MUST, highlighting its broad discovery potential and the synergy it offers with other leading facilities.

\subsubsection{Cosmic Expansion and Dark Energy}
\label{ssec:dark-energy}

    By the late 2020s, Stage-IV projects such as DESI and Euclid will have measured the cosmic expansion history up to $z \sim 2$ with sub-percent precision using baryon acoustic oscillations (BAO) as a standard ruler \cite{2016arXiv161100036D, 2025A&A...697A...1E}. Notably, recent DESI data suggest that the equation of state of dark energy may vary with time \cite{2025arXiv250314738D}. This underscores the need for more powerful surveys to unravel the nature of dark energy. MUST will extend cosmic expansion measurements well beyond $z \sim 2$, achieving percent-level precision with high-redshift tracers such as Lyman-break galaxies (LBGs) and Lyman-$\alpha$ emitters (LAEs) \cite{2024arXiv241107970Z}. Consequently, MUST will probe the time evolution of dark energy with unprecedented precision up to $z \sim 5$, enabling either the detection of deviations from $\Lambda$CDM (such as early dark energy or other exotic components), or robustly confirming that dark energy behaves as a cosmological constant.

    Furthermore, MUST will complement other probes of dark energy. For instance, it will provide extensive spectroscopic calibration samples for imaging surveys like Euclid and LSST. This will enhance dark energy constraints by improving the accuracy of lensing tomography and cluster cosmology. Besides, by collaborating with gravitational-wave (GW) detectors, MUST will enable new methods such as standard sirens \cite{1986Natur.323..310S, 2017Natur.551...85A, 2025arXiv250200239P}. Combining GW observations of merging neutron stars with precise redshift measurements will yield $H_0$ constraints from luminosity distances that are largely independent of both CMB data and the distance ladder, helping to resolve the Hubble tension.

\subsubsection{Structure Growth and Gravity}
\label{ssec:modified-gravity}

Besides cosmic expansion, MUST will trace the growth of structure through the redshift-space distortion (RSD) effect and provide stringent tests of gravity on cosmic scales. Notably, MUST will deliver percent-level measurements of structure growth up to $z \sim 5$, deep in the matter-dominated era  \cite{2024arXiv241107970Z}. The combination of growth rate and expansion history measurements will allow detections of mild deviations from $\Lambda$CDM, such as a breakdown of General Relativity (GR) or interactions between dark energy and matter.

Cross-correlating MUST tracers with lensing maps from CMB-S4 and imaging surveys will further enhance gravity tests. This synergy breaks parameter degeneracies and enables precise measurements of matter clustering amplitude that cannot be obtained by single probes alone. For instance, combining galaxy peculiar velocities and weak lensing can provide a model-independent way to constrain modified gravity parameters \cite{2011PhRvD..84h3523S}. These analyses can help clarify current tensions between lensing- and CMB-inferred clustering amplitudes.

\subsubsection{Neutrino Physics and Light Relics}
\label{ssec:neutrino}

Though neutrinos are nearly massless and contribute only a small fraction of matter density, they leave a cumulative gravitational effect on the clustering of cosmic structures. As neutrinos stream out of over-dense regions, they suppress the growth of structure and damp small-scale clustering. This scale-dependent suppression encodes information about the total neutrino mass ($\sum m_\nu$) \cite{1998PhRvL..80.5255H, 2006PhR...429..307L}. Currently, the tightest constraint ($\sum m_\nu < 0.072\,{\rm eV}$ at 95\% CL) comes from the combination of CMB and DESI data \cite{2025JCAP...02..021A}, approaching the lower bound from the normal mass hierarchy (0.06\,eV). In contrast, the lower limit of the inverted hierarchy is $\sim 0.1\,{\rm eV}$. Distinguishing between the two hierarchies will provide direct insights into whether neutrinos are Dirac or Majorana particles \cite{2015PrPNP..83....1Q}. MUST will be pivotal in this effort: by mapping structures at high density, it will substantially reduce the statistical uncertainty of $\sum m_\nu$, particularly when combined with future CMB data.

Beyond the known neutrinos, cosmological surveys can also probe additional relativistic species of `light relics' in the early Universe, characterized by the effective number of neutrino-like species $N_{\rm eff}$. Extra light particles such as sterile neutrinos and axions would alter the expansion rate and introduce phase shifts in the BAO signal \cite{2017JCAP...11..007B, 2018JCAP...08..029B, 2019NatPh..15..465B, 2025MNRAS.538.1980W}. By extending BAO measurements to higher redshifts, MUST will help break degeneracies in CMB-only constraints and search for signatures of extra radiation in the primordial plasma. Precise measurements of $\sum m_\nu$ and $N_{\rm eff}$ with MUST will deepen our understanding of the neutrino sector and potentially reveal new physics beyond the Standard Model, such as non-standard neutrino interactions or other light relic particles.

\subsubsection{Inflation and Primordial Physics}
\label{ssec:primordial-inflation}

The large volume of galaxies mapped by MUST will permit precise clustering measurements on very large scales that encode signatures of primordial physics. One key example is the detection of primordial non-Gaussianity (PNG), a powerful probe of the mechanism and energy scale of cosmic inflation \cite{2005PhRvL..95l1302L, 2010AdAst2010E..89D}. MUST is expected to measure the local-type PNG parameter ($f_{\rm NL}^{\rm local}$) at the $\mathcal{O} (1)$ level, surpassing the current CMB constraints \cite{2024arXiv241107970Z}. This precision would allow discrimination between single-field slow-roll inflation (which predicts $f_{\rm NL}^{\rm local} \ll 1$) and alternative models involving multiple fields, non-equilibrium dynamics, or high-energy particle contents, many of which predict $f_{\rm NL}^{\rm local} \gtrsim 1$ \cite{2006PhRvD..73h3522R, 2011JCAP...09..014B, 2011PhRvD..84l3506K, 2012PhRvD..85b3531A}.

The high density of MUST tracers will also support searches for PNG in higher-order statistics (such as the 3- and 4-point correlation function) and for exotic correlation patterns beyond the traditional $f_{\rm NL}$ shapes. For example, the `cosmological collider' framework suggests that heavy particles present during inflation could leave oscillatory imprints in the clustering of large-scale structures (LSS) that encode their mass and spin \cite{2015arXiv150308043A, 2019JHEP...12..107K}. Additionally, detecting parity-violating patterns in LSS would provide clear evidence of new physics beyond the standard inflation paradigm. Thus, MUST will provide stringent tests of high-energy physics in the primordial Universe.

\subsubsection{Dark Matter}

    The nature of dark matter remains one of the most significant open questions in astrophysics and particle physics. While the cold dark matter (CDM) paradigm is widely accepted, various alternatives -- such as warm dark matter, wave-like dark matter, and self-interacting dark matter -- have been proposed to address the lack of direct detection and the small-scale challenges to CDM, including the `missing satellite' and cusp-core problems \cite{2001ApJ...556...93B, 2015PNAS..11212249W, 2015MNRAS.451.2479M, 2018PhR...730....1T}. MUST will offer several means to distinguish among these models. One powerful approach involves detecting the small-scale cutoff in the clustering of Lyman-$\alpha$ forests, measured from high-redshift tracers such as quasars and LBGs \cite{2017PhLB..773..258G, 2017MNRAS.471.4606A, 2021PhRvL.126g1302R}. The unprecedented density of these tracers will enable tight constraints on dark matter scenarios. Another method focuses on detailed mapping of the Milky Way and Local Group halos. MUST will measure radial velocities and chemical abundances for millions of halo stars. This will enable the reconstruction of the dynamics and density profiles of these halos, providing tests of dark matter properties. In short, MUST will integrate advances across astronomy, cosmology, and fundamental physics to understand the nature of dark matter.

\subsubsection{Galaxy Formation and Evolution}

    While focusing on the large-scale structure, MUST will also bring the statistical studies of galaxies and Active Galactic Nuclei (AGN) into a new era. Firstly, by leveraging the spectra of hundreds of millions of galaxies for the Stage-V cosmology survey, MUST will provide critical statistical insights into the formation and evolution of galaxies across cosmic time. At high redshifts ($z \gtrsim 2$), observations of LBGs, LAEs, and QSOs will probe the early stages of galaxy assembly, including the mechanisms of star formation \& quenching, chemical enrichment, and gas accretion in young galaxies. These measurements will also help constrain models of cosmic reionization. In the low-redshift Universe ($z \lesssim 1.5$), MUST will enable detailed studies of galaxy properties across different environments -- from isolated field galaxies to dense clusters -- examining the complex interactions between galaxy evolution and their surrounding ecosystems.

    In addition to the statistical studies of extragalactic targets piggybacked on the cosmology survey, MUST's excellent multiplexed survey capability enables it to conduct more specialized extragalactic surveys in the bright night sky or design ancillary programs to observe specific populations of galaxies. For instance, at $z<0.2$, MUST could target star-forming dwarf galaxies down to $M_{\star}\sim 10^{8}\ \mathrm{M}_{\odot}$, which could dramatically deepen our understanding of the galaxy-halo connection of dwarf galaxies with the help of the galaxy-galaxy lensing data from the same field. Also, strategically targeting the opposite ends of low-redshift disk galaxies could help infer their rotational velocities, which could be used to support an extensive peculiar velocity survey that provides complementary cosmological information (e.g., \cite{2023MNRAS.525.1106S}). Like these examples, there could be many other possibilities for MUST to become an essential facility to advance our knowledge about galaxy formation and evolution in the 2030s.

\subsubsection{Milky Way Structure and Galactic Archaeology}

    Carrying the legacy of the LAMOST, DESI, and other low-to-mid resolution surveys of Milky Way stars, MUST is expected to obtain spectra of tens of millions of stars across various Galactic components, mostly during bright and gray time. These observations will provide a detailed map of the Galactic halo and outer disk. Combined with the astrometric data from missions like {\it Gaia}, the resulting 6-D phase-space information (positions and velocities) will help reconstruct the assembly history of the Milky Way by identifying stellar streams and substructures. Compared to the Milky Way surveys of DESI \cite{2023ApJ...947...37C} and 4MOST \cite{4MOST_CS1, 4MOST_CS2}, MUST can reach fainter and more distant stellar populations in the stellar halo while achieving similar precision for radial velocity, which is crucial for characterizing Milky Way's recent assembly and inferring its dark matter halo properties (e.g., mass, 3-D shape). And, by measuring the chemical compositions of different stellar populations, MUST will help trace the chemical enrichment history of the Milky Way and identify stars from the same progenitor systems. In short, this deep and large stellar sample will provide powerful probes of Galactic dynamics and early star formation.

    At the same time, an extensive stellar survey in the cosmological field will help provide a detailed map of Galactic extinction, which is identified as a potentially critical systematic issue for a Stage-V spectroscopic survey. Therefore, the MUST's Milky Way survey will also be a vital component to support its primary scientific goals.

\subsubsection{Time-domain and Multi-messenger Astrophysics}

    The wide field, large aperture, and multiplexed fiber system of MUST will enable rapid spectroscopy of transient targets, even for those that are relatively faint or in crowded fields. MUST will support flexible survey scheduling by allocating a subset of fibers for prompt transient follow-up alongside regular survey exposures. By efficiently classifying large numbers of transients, MUST will build extensive samples for statistical studies of event subtypes, physical mechanisms, and environmental dependencies for various transient events, including supernovae and fast radio bursts. Meanwhile, MUST will systematically monitor transient sources, including variable stars, tidal disruption events, and quasars.

Multi-messenger sources are also a key target of MUST. For instance, spectroscopic follow-up of high-energy astrophysical accelerators, such as supernova remnants and blazars, will provide deeper insights into the emission mechanisms. In addition, by measuring redshifts for the hosts of GW events, MUST will enable the use of standard sirens as an independent cosmic distance probe. In short, MUST will provide a much richer view of the dynamic Universe, from fast explosive transients to long-period variables.

\begin{figure*}
\centering 
\includegraphics[width=0.69\textwidth]{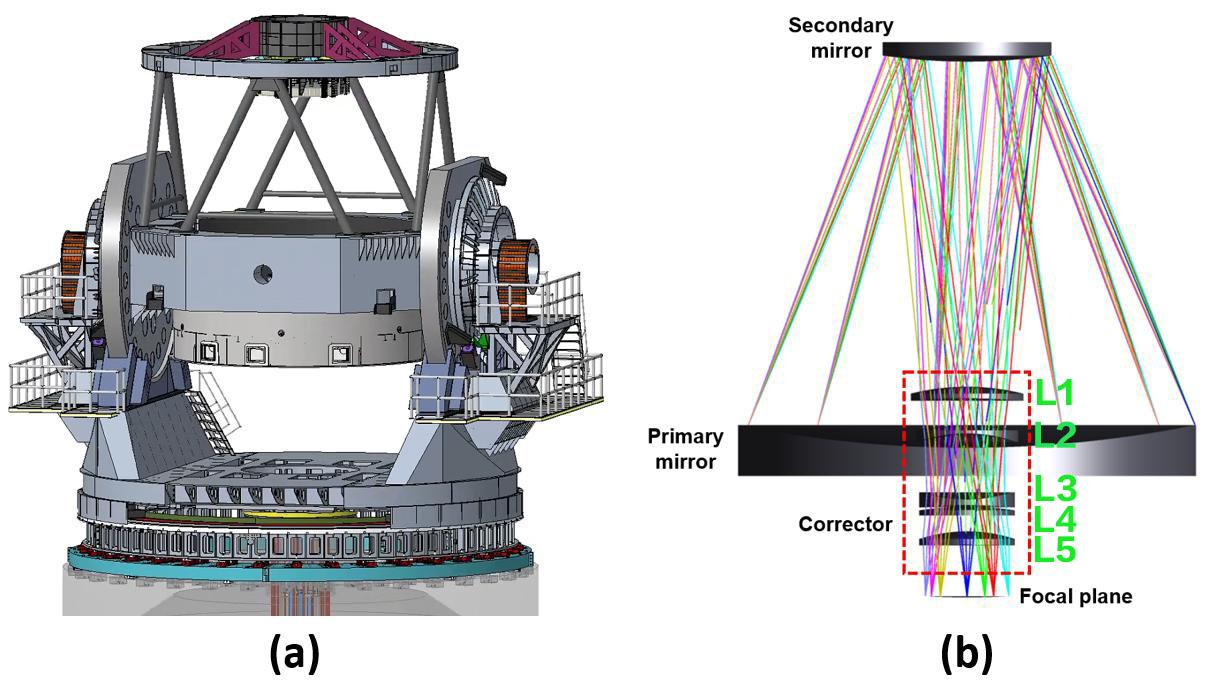}
% \caption{(a) Schematic of the 6.5\,m MUltiplexed Survey Telescope (MUST); (b) Optical design of MUST.}
\caption{
    The MUST schematic (a) and the corresponding optical design (b). MUST is a reflective telescope that adopts a Ritchey–Chretien (R-C) + Wide-Field Corrector (WFC) design, featuring a 6.5-m concave hyperboloid primary mirror and a 2.4-m convex hyperboloid secondary mirror. Upon completion, both mirrors will be the largest of their kind ever manufactured in China. The WFC system (cf. Figure \ref{fig:WFC_system}) is rigidly linked to the primary mirror system for delivering high imaging quality across a wide field of view.}
\label{fig:Schematic_Optical_design}
\end{figure*}

\subsection{MUST: a Stage-V Spectroscopic Facility}
    \label{ssec:must-key-prop}

    From SDSS to DESI, from Stage-II to Stage-IV spectroscopic surveys, every new stage was marked by a significant increase in the telescope's light-collecting power, the focal plane's multiplex capabilities, and the overall survey efficiency. As we enter the era of Stage-V spectroscopic survey for cosmology, we face new requirements and challenges in designing the telescope, its scientific instruments, and the supporting facilities. MUST is on its way to becoming the first operating Stage-V facility. Although its original design was greatly inspired by the MegaMapper concept \cite{2019BAAS...51g.229S}, the detailed optical and engineering design has gradually matured through exploration and experiments led by Tsinghua University and other domestic and international collaborating institutes.

    Here, we provide a summary of the current design concept for MUST, along with the engineering challenges it faces during the design, construction, and operational phases. While some of these challenges are unique to MUST, many of them will be common among other Stage-V and future spectroscopic facilities. 
    
    It is worth reminding readers that, as we prepare for this review, the MUST telescope has entered the detailed design and early production phase; however, many technical details are still subject to modification and improvement. Meanwhile, the summit facility for MUST, the focal plane system, and the spectrographs are still in the preliminary design stage. We will provide detailed technical summaries for these systems after their designs are finalized.

\subsubsection{Overview of the MUST Facility}
    \label{ssec:overall-design}

\begin{figure*}
    \centering 
    \includegraphics[width=0.7\textwidth]{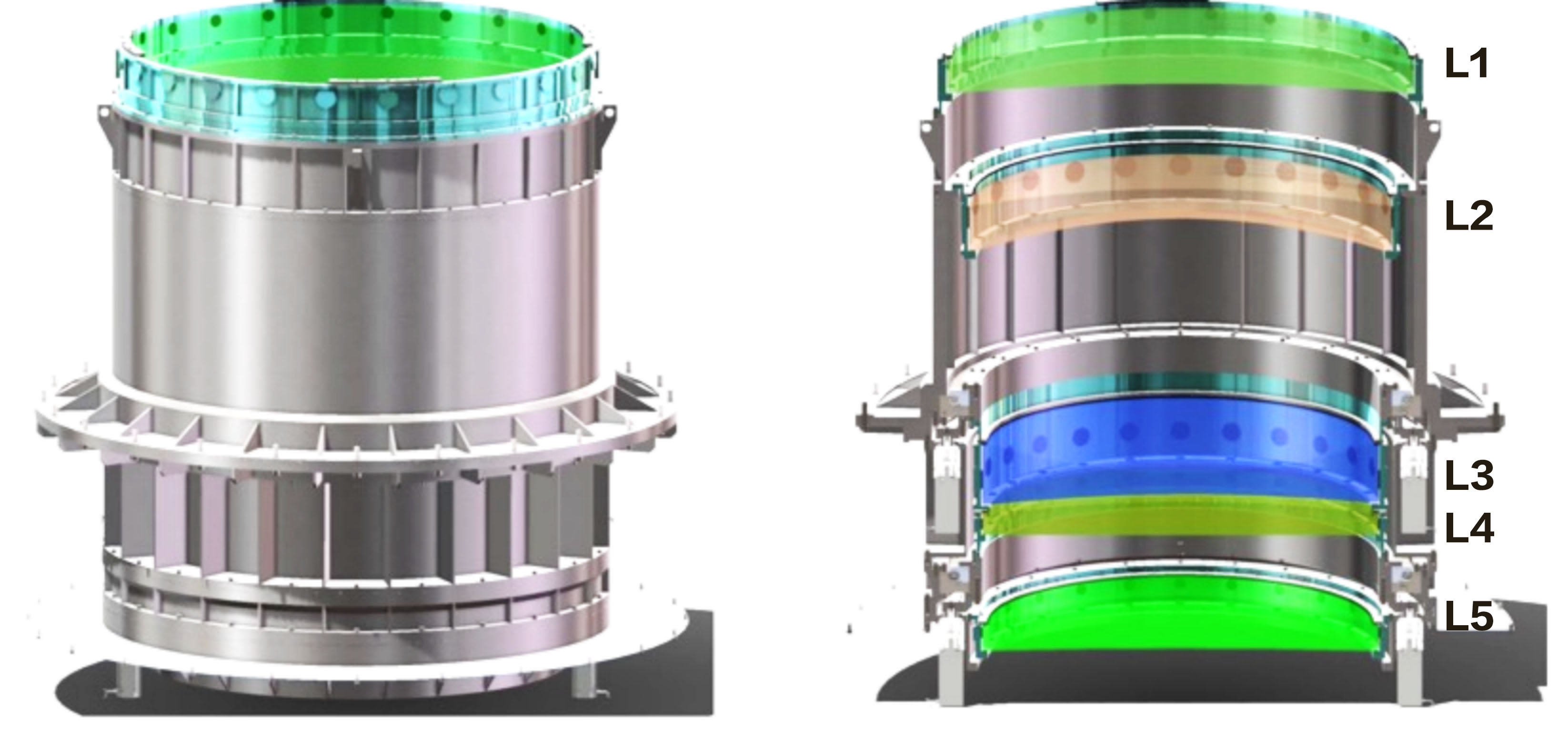}
    \caption{The schematic of the MUST WFC system. The WFC consists of lenses L1 $\sim$ L5 as labeled in the right panel, with aspheric surfaces on L1, L2, and L5. The WFC has a length of approximately 2.3 meters and a total mass of around 7 tons.}
    \label{fig:WFC_system}
\end{figure*}

    As a spectroscopic survey facility designed for the 2030s and beyond, MUST is much more than just a telescope. Firstly, MUST is the facility responsible for conducting a major experiment for fundamental physics - the Stage-V cosmology survey. Equally importantly, MUST will be a powerful and flexible spectroscopic platform for the Chinese and international astronomical communities.

    Located at the 4,380m Peak A of the Saishiteng Mountain in Lenghu, the high altitude is the first critical challenge for MUST. MUST's summit facility, including the telescope dome (cf. Figure\,\ref{fig:must_overview}) and the supporting infrastructure, will be carefully designed to endure the harsh environment while improving the local observing conditions, such as the dome seeing. The MUST telescope and the instruments shall be able to observe when the temperature is higher than -20\ $^{\circ}$C and survive (no permanent function loss) even more extreme conditions at $>-40\ ^{\circ}\rm C$. In addition to the telescope dome, the MUST summit facility will also include a special environment-controlled room for the spectrograph systems and other instruments, a large workshop to support the on-site maintenance of the telescope (e.g., recoating of the primary mirror), and a clean room for maintaining the spectrograph. 

    At the center of the MUST project is a 6.5-m wide-field telescope for spectroscopic survey. Unlike the Stage-IV spectroscopic facilities, such as DESI and PFS, which modified existing telescopes (the 4-m Mayall and 8.2-m Subaru telescopes) to host fiber positioning systems at the primary focus, MUST is a new telescope designed explicitly for wide-field spectroscopic surveys. MUST adopts the Ritchey–Chretien (R-C) $+$ Wide-Field Corrector (WFC) design, which has been proven to be a practical approach for achieving high imaging quality over a large field of view. As shown in Figure\,\ref{fig:must_overview} and Figure\,\ref{fig:Schematic_Optical_design}, MUST's focal plane is located at the Cassegrain focus, which makes it easier to host a massively multiplexed fiber positioning system that would be too large and heavy for the primary focus. A Cassegrain focal plane also makes it easier to maintain and upgrade the instrument. Moreover, as MUST and future Stage-V facilities will host $>$10,000 fiber positioners, the arrangement and movement control of the fiber cable route will become a major challenge. We anticipate that the choice of Cassegrain focus can also help reduce the difficulty of feeding photons from the tip of the positioners to the spectrograph without incurring significant throughput loss or excessive additional focal ratio degradation (FRD). To support the entire optical and focal plane system, MUST's mount and control system will be designed to satisfy the pointing and tracking requirements.

    Under the current design, as shown in Figure\,\ref{fig:must_overview}, the fiber cables will pass through the central hole below the Cassegrain focus to a circular spectrograph room that will be temperature- and humidity-controlled during the operation. MUST plan to host roughly 40 three-channel multi-object spectrographs that cover the wavelength range between 370 to 960 nm with low-to-mid resolution ($R=\lambda/\mathrm{d}\lambda \sim 1800$) to 5000. Given the key scientific goals and potential targets for a Stage-V cosmological survey, the key specifications of the spectrograph satisfy the scientific requirements and enable us to use the mature designs for Stage-IV experiments (e.g., DESI) as a reference.

    We will separately provide more details about MUST's optical design, focal plane system, and spectrograph systems in the next few subsections. Meanwhile, it is worth mentioning that, to facilitate the scientific operation and ensure the scientific productivity, MUST also requires a sophisticated data system and computing infrastructure to not only oversee the transfer, storage, and back-up of different scientific and engineering data, but also support the demanding requirements for data reduction and cosmological analysis.

\begin{table*}[htbp]
\centering
\caption{The MUST Optical System Parameters and Requirements.}
\resizebox{0.6\linewidth}{!}{
\begin{tabular}{|c|c|}
\hline
\textbf{Parameter} & \textbf{Requirement} \\ \hline
Optical structure & Ritchey-Chretien with Wide Field Corrector (WFC) \\ \hline
Focus type & Cassegrain focus \\ \hline
Focal length & 24088 mm \\ \hline
Wavelength band & 360 nm to 1000 nm \\ \hline
Entry diameter & 6.5 m \\ \hline
Field of view & $\pm 1.4^\circ$ \\ \hline
Focal ratio & F/3.717 \\ \hline
EE80 size of image spot & $<$ 0.52 arcsec in diameter \\ \hline
Focal plane size & 1.2 m \\ \hline
\end{tabular}}
\label{tab:must_parameters_requirement}
\end{table*}

\subsubsection{The Optical and Mount System}
    \label{ssec:overview-TOS}

    In Figure\,\ref{fig:Schematic_Optical_design}, we illustrate the overall optical design of MUST. As mentioned earlier, MUST is a reflective telescope that adopts a popular R-C design. It has a 6.5-m concave hyperboloid primary mirror and a 2.4-m convex hyperboloid secondary mirror. Upon completion, both of the primary and secondary mirrors will become the largest of their kind ever manufactured in China. 
    
    In addition to the mirror itself, the primary mirror system also consists of a mirror chamber and an active support system \cite{Zhang2023ConceptualDO, 2024arXiv241107970Z, 2025PASP..137b5001Z}. The lightweight primary mirror relies on a hydraulic active support system and multiple wavefront sensors located on the focal plane to enable real-time correction of the mirror surface to an expected root mean square (RMS) value of less than 15 nm. In addition to the high imaging quality, MUST's optical design also prioritizes the compactness of the telescope itself. To shorten the optical path and reduce the size of the secondary mirror, the primary mirror was designed with a fast focal ratio of F/1.25. 

    As for the secondary mirror, the lightweight structure design, the precise polishing, and the accurate examination of this large 2.4-m convex mirror present several critical challenges, which have been carefully investigated. Currently, the whole secondary mirror system, including the axial \& radial support and the mirror chamber, weighs less than 1 ton. MUST's secondary mirror is passively supported, utilizing a 36-point flexible Whiffle-tree mechanism for axial support and a 15-point astatic lever structure for radial support. Altogether, they can achieve a surface shape accuracy of $<25$ nm RMS. The lightweight process and other early production processes have begun for the secondary mirror system.

    Meanwhile, on top of the R-C design, an ambitious five-lens WFC is the key to MUST's excellent imaging quality over the desired wide FoV and is fundamental to achieving MUST's scientific goals. The WFC system will correct the image quality within a circular FoV, whose diameter is $\sim 2.8^{\circ}$, corresponding to a 1.2-m focal plane. As shown in Figure\,\ref{fig:WFC_system}, the WFC system includes five large-aperture lenses, their own lens cells, and a corrector barrel. Among these lenses, L1, L2, and L5 have aspheric surfaces, which makes the manufacturing and examination processes more challenging. In particular, the L1 lens has a diameter of 1.6 m, making it the largest aspheric lens in the world, even larger than the largest lens for the Vera Rubin Observatory's wide-field corrector. Meanwhile, L3 and L4 are tilted lenses, acting as atmospheric dispersion correctors (ADCs) to support observations within the zenith angle range of 0 to 60 degrees. The WFC system also features a five-degree-of-freedom adjustment mechanism that is rigidly connected to the primary mirror system. The total length of the WFC is approximately 2.3 meters, and its total mass is approximately 7 tons. We should emphasize that MUST's WFC system is an extremely challenging system, whose design must balance the achieved imaging quality with the feasibility of manufacture and validation. The final design has meticulously taken the size \& quality of the available glass materials, the achievable fabrication precision of the aspheric lens, and the tolerance budget. We will present the detailed design of the WFC system in future publications. The procurement of glass blanks for all five lenses has been completed, and the fabrication process has commenced.

\begin{figure*}
    \centering 
    \includegraphics[width=0.7\textwidth]{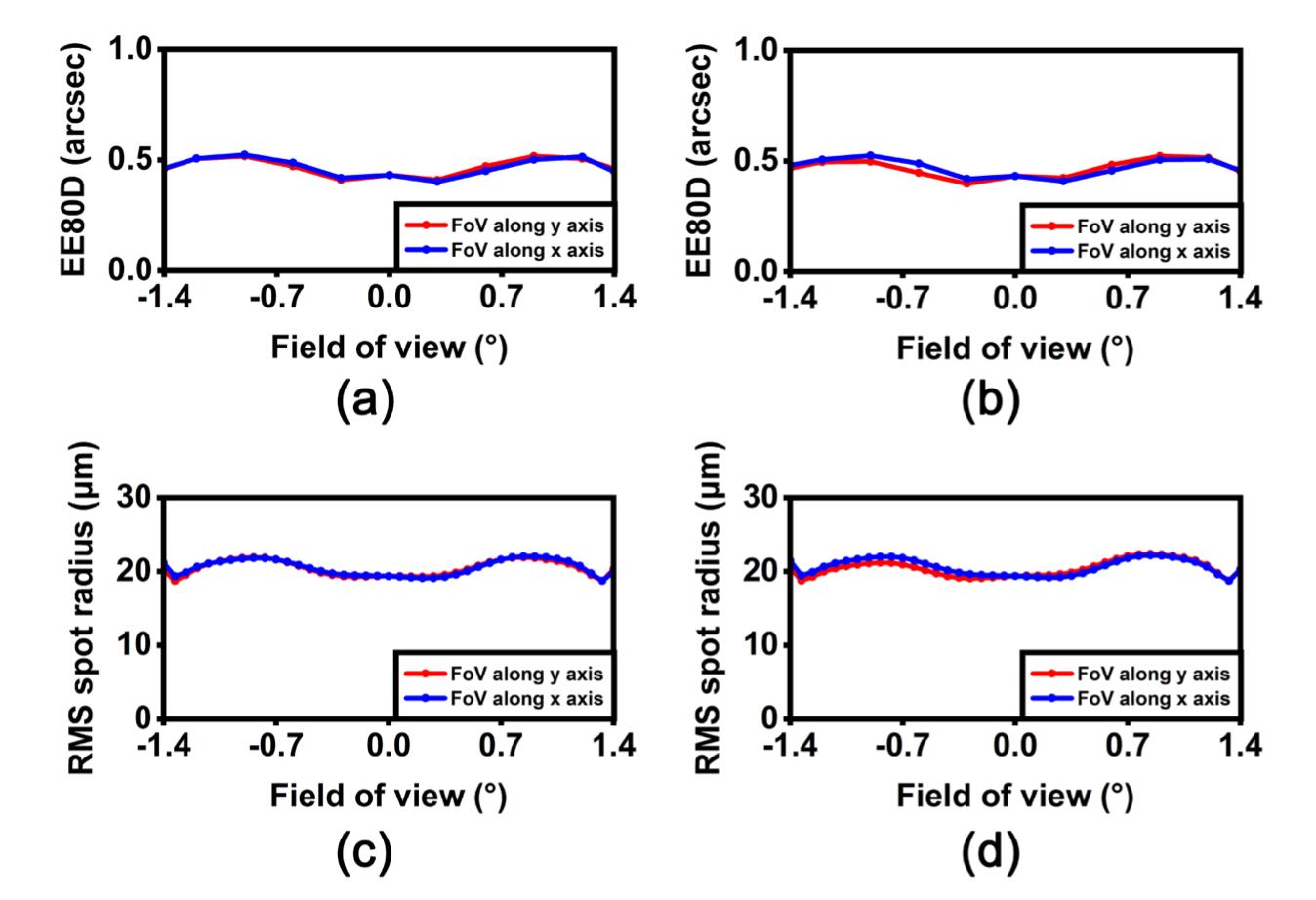}
    \caption{The EE$80$ over 2.8$^\circ$ FoV of MUST at (a) 0$^\circ$ and (b) 55$^\circ$ zenith angle. The RMS spot radius over 2.8$^\circ$ FoV of MUST at (c) 0$^\circ$ and (d) 55$^\circ$ zenith angle. The high and homogeneous imaging quality satisfies all of MUST’s performance requirements. As shown, the EE$80$ size of image spots is under 0.52$''$, demonstrating high and homogeneous imaging quality that meets all of MUST's performance requirements.}
\label{fig:EE80_RMS}
\end{figure*}

    Detailed optical analysis confirms that this design can achieve excellent imaging quality over the whole FoV, within the required range of zenith angle, and across the entire observing wavelength range for MUST (370-960 nm), with the 80\% Encircled Energy (EE80) size of image spots less than 0.52$''$ (cf.\,Figure\,\ref{fig:EE80_RMS}) \cite{Zhang2023ConceptualDO, 2024arXiv241107970Z, 2025PASP..137b5001Z}. The high and homogeneous imaging quality satisfies all of MUST's performance requirements.

    With this optical design, MUST's total focal length is 24,088 mm, resulting in a $\sim$ F/$3.71$ system focal ratio, which is faster than most 6-8 meter telescopes at the Cassegrain focus. At the same time, to control the field curvature over the FoV, MUST's focal plane is designed to be a 6th-order aspheric surface. Compared to a spherical focal surface, it slightly increases the difficulty in developing the focal plane system.

    To operate such an ambitious and precise optical system, MUST's mount and control system is designed to achieve the accurate pointing and continuous tracking requirements for a Stage-V cosmological survey. MUST has finished the preliminary design of the elevation axis, azimuth axis, triangular support, azimuth platform, and leveling mechanism of the mount system \cite{Zhang2023ConceptualDO}. Both the elevation and azimuth axes utilize hydraulic support and a direct drive configuration with spliced torque motors to achieve a pointing accuracy of less than 4$''$, an open-loop tracking accuracy of less than 0.2$''$, and a guided star closed-loop tracking accuracy of less than 0.1$''$.

    We summarize the primary specifications of the 6.5-m telescope for MUST in Table \ref{tab:must_parameters_requirement}.
    
\subsubsection{Focal Plane and Fiber System}
    \label{ssec:fps-design}

    MUST's focal plane can be described as a circular region with a diameter of 1,189.306 mm and image scale of 116.784 $\mu\rm{m}/''$. Its overall shape is an aspheric surface that the following equation can describe:

    \begin{equation}
        Z = \frac{c_s r^2}{1 + \sqrt{1 - (1 + k_s){c_s}^2 r^2}} + {A_2}r^2 + {A_4}r^4 + {A_6}r^6 
    \end{equation}

    \noindent{}where $r$ is the distance to the center of the focal plane, $c_s=1/11698.7\rm mm$, $k_s=0$, $A_2=0/\rm mm$, $A_4=3.76762\times 10^{-11}/\rm mm^3$, and $A_6=-5.86796\times 10^{-17}/\rm mm^5$. 

    The critical task for MUST's focal plane system is to host more than 20,000 fiber positioners and ensure they can be reliably and repeatedly reconfigured to collect photons from the desired targets without inducing excessive additional light loss or geometric FRD, which would slow down MUST's survey speed. While fiber positioners have been introduced to spectroscopic observation for almost 30 years, and an automatic fiber positioning technique has existed since the 2dF days, the ambitious scientific goals of the Stage-V cosmology put many stringent technical requirements on the focal plane system. 

    In particular, after each reconfiguration, the offset between the final position of each positioning robot's fiber center and the commanded position should have a $\leq 6\ \mu\rm{m}$ RMS when projected to the X-Y plane that is perpendicular to the optical axis of the telescope. At the same time, the deviation between the final position of each fiber's upper surface and the ideal aspheric focal surface should be within $\pm\ 50\ \mu\rm{m}$ when projected to the Z-axis (along the optical axis). The tilt angle between each positioning robot's fiber tip and the telescope's incoming chief ray shall be $\leq 0.3^{\circ}$. Such a challenging positioning requirement can only be achieved with an innovative design of the positioner system, a group of back-illuminating fiducial fibers, and a metrology camera system at the center of the secondary mirror. 

\begin{table*}
    \centering
    \caption{The specification of the MUST spectrograph. MUST will equip itself with 40 spectrographs, taking into account performance, risk, and cost. This setup can ensure excellent throughput, vital for meeting Stage-V survey sensitivity and efficiency requirements, while also facilitating consistent camera alignment across all 40 spectrographs.}
\resizebox{0.6\linewidth}{!}{
\begin{tabular}{|c|c|c|}
\hline 
& \textbf{Item} & \textbf{Specification} \\
\hline 
\multirow{5}{*}{Performance} & Bandpass & 370 nm $\sim$ 960 nm \\
\cline{2-3} 
& Spectral resolution & \begin{tabular}{@{}l@{}}
360 nm $\sim$ 590 nm: $>$1,500 \\
565 nm $\sim$ 775 nm: $>$3,000 \\
750 nm $\sim$ 960 nm: $>$4,000
\end{tabular} \\
\cline{2-3} 
& Throughput & Peak $>$ 50\% for each channel \\
\hline 
\multirow{5}{*}{Parameters} 
& Fiber core diameter & 140 $\mu$m \\
\cline{2-3} 
& Multiplexing & 500 fibers \\
\cline{2-3} 
& Detector pixel pitch & 15 $\mu$m \\
\cline{2-3} 
& Detector element number & $>$4,000 \\
\cline{2-3} 
& Sampling pixel number & 4 \\
\hline 
\multirow{4}{*}{Environment} 
& Operational temperature & 20$^{\circ}$C $\pm$ 2$^{\circ}$C \\
\cline{2-3} 
& Survival temperature & -30$^{\circ}$C $\sim$ 30$^{\circ}$C \\
\cline{2-3} 
& Operational humidity & 0 $\sim$ 85\% RH \\
\cline{2-3} 
& Pressure & 0.6 $\sim$ 1.0 atm \\
\hline
\end{tabular}
}
\label{tab:must_spectrograph}
\end{table*}

    To meet these demanding requirements, MUST is collaborating with EPFL and other industrial partners to develop a novel modular focal plane system \cite{2024arXiv240816596R} that was first proposed in the context of the MegaMapper concept \cite{2019BAAS...51g.229S}. As shown in Figure\ \ref{fig:must_overview}, we plan to host 336 triangular fiber positioner modules on a 1.2-m diameter stainless steel focal plate. Each module will integrate 60 fiber positioning robots in three groups and three reference surfaces to facilitate the precise installation of the module to the focal plate. All the modules are identical and interchangeable, which means that the tips of the positioners exhibit systematic deviation from the ideal aspheric focal surface within each module and across the entire focal plane. The position \& direction of each module, along with the dimension of the module, are carefully designed to ensure the defocus requirement is met. 

    MUST plans to adopt a miniaturized fiber positioning robot with an outer diameter of 6.2 mm and an optical fiber with a 140 $\mu\rm{m}$ core diameter, corresponding to 1.2 arcseconds on the sky, to ensure the fiber density meets the requirements of future cosmology surveys. To optimize the effective coverage of the fibers, the focal plate assembles four modules into a group. The gaps between modules are 1 mm and 3 mm within the group and between adjacent groups. Altogether, this allows MUST to equip 21,168 fiber positioners over the focal plane, achieving a 74\% coverage. 

    This novel and challenging modular design not only satisfies the requirements of a Stage-V cosmological survey but also secures future opportunities for replacing and upgrading focal plane instruments. Currently, the design of the modular focal plane system and the fiber positioning robot is in the preliminary phase. Meanwhile, we are conducting the conceptual design for the rest of the complicated focal plane system, which includes, but is not limited to:

    \begin{itemize}

        \item The hexapod \& derotator system, which serves as the physical interface with the primary mirror system. It will also correct for field rotation during the observation, while consistently ensuring the precise alignment of the focal plane system with the telescope's optical system. 
        
        \item The enclosure assembly to provide physical protection, electronic control, and cooling \& ventilation for the focal plane system. 

        \item The guiding cameras and the wavefront sensors located on the focal plate that ensure the precise pointing \& tracking of the telescope and provide feedback to the active optic system. In addition, the guiding cameras and the fiducial fibers on them define the reference for the focal plane's coordinate system. 
        
        \item Scientific fiber and fiber cable system that carefully pass around 20,000 delicate fibers from the positioning module to the spectrograph through a $\sim 45$ meters long fiber cable bundles. During this process, the fiber route needs to account for any rotational or other movements of the cable while inducing as little additional tension as possible on the fiber to control the FRD budget. 

    \end{itemize}

\subsubsection{Spectrograph System}
    \label{ssec:spec-design}

\begin{figure*}
    \centering 
    \includegraphics[width=0.85\textwidth]{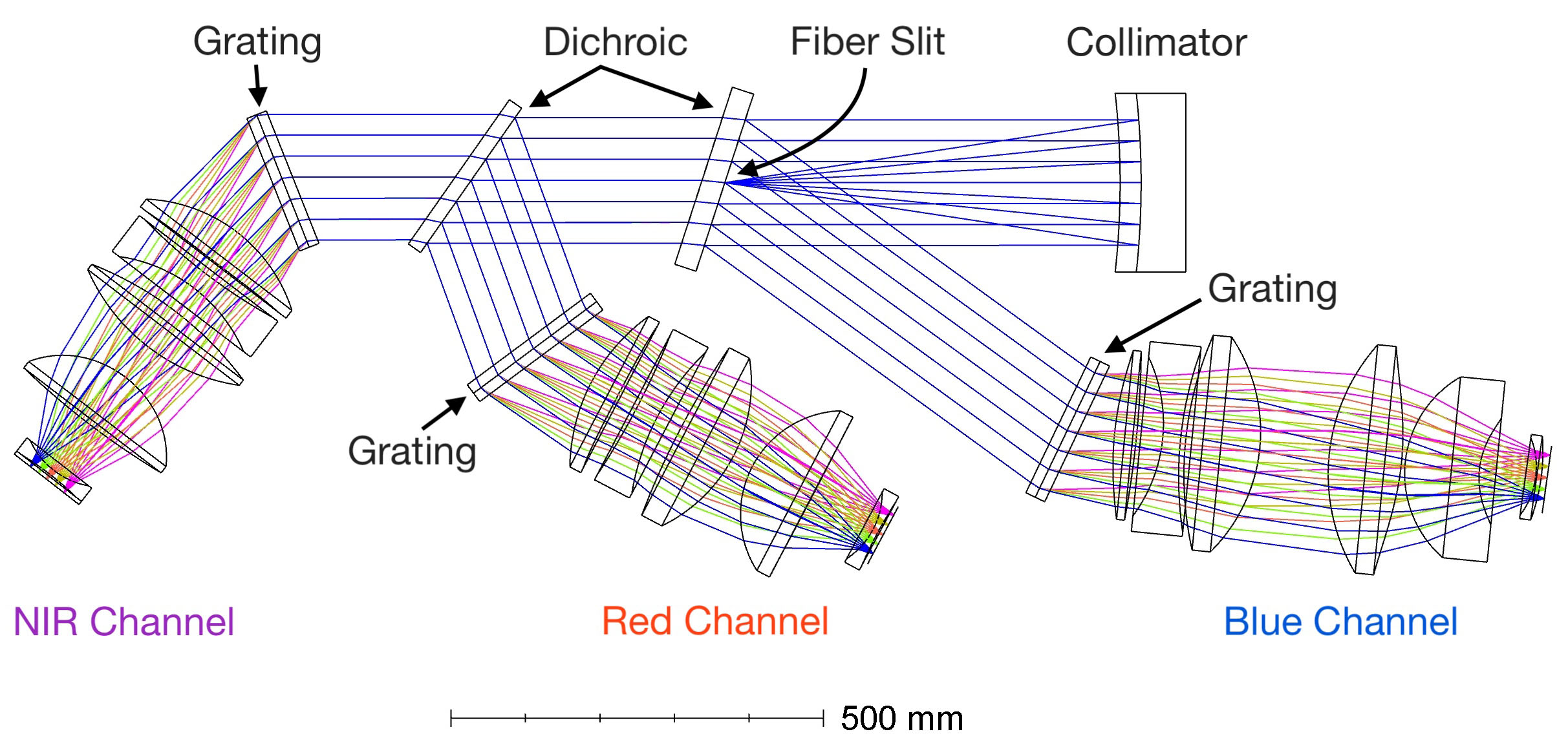}
    \caption{The optical layout of one candidate design of MUST's spectrograph. Similar to DESI, this design features an on-axis fiber slit (located inside the blue dichroic), a reflective spherical collimator, three transmissive gratings, and a five-lens transmissive camera design for all three channels. The blue channel is isolated first to improve its overall throughput, accommodating the critical requirements for high-redshift cosmology, as the Lyman-$\alpha$ emission line of a $2<z<5$ LBG or LAE falls within the wavelength range of the Blue channel. The last lens of each channel will be a field lens that is integrated into the cryostat.}
\label{fig:spectrograph}
\end{figure*}

    As mentioned above, the wavelength coverage and the spectral resolution of the spectrograph for the Stage-IV experiment (e.g., DESI) can also facilitate the scientific requirements of a Stage-V survey. However, while the conceptual design of MUST's spectrograph system can learn a great deal from the DESI one, in practice, it still faces numerous significant challenges. In particular, after taking the FRD budget of the fiber system and the fiber pseudo-slit into account, MUST's spectrograph needs to accommodate a F/3.3 input beam, much faster than DESI's. This choice significantly increases the difficulty of camera design. Moreover, compared to DESI, MUST's spectrograph system will support about 20,000 fibers, which will substantially increase the number of required spectrographs. Therefore, the spectrograph design must consider more factors than just spectral performance, such as cost, fabrication difficulty, assembly, alignment, operation, and maintenance of the entire spectrograph system. Moreover, several practical constraints also limit the feasible designs for MUST. For instance, the availability of high-performance CCDs, transmissive gratings, and specific glass materials in China poses a significant risks that limit the design.

    After carefully balancing the performance, risk, and cost of different designs, we chose to support MUST with 40 spectrographs, each hosting $\sim 500$ 140 $\mu$m fibers on a 60\,mm $\times$ 60\,mm detectors (4k $\times$ 4k pixels, with a 15 $\mu$m pixel size). Each spectrograph has three channels that cover the following ranges: 360-590 nm (Blue), 565-775 nm (Red), and 750-960 nm (NIR). We adopt the reflective collimator and transmissive camera design used in DESI. Such a combination can achieve excellent throughput, which is crucial for meeting the sensitivity and efficiency requirements of a Stage-V survey, while also facilitating consistent alignment of the cameras for 40 spectrographs. We summarize the essential design requirements for MUST's spectrograph in Table\,\ref{tab:must_spectrograph}.

    As the spectrograph design is still in the preliminary phase, we illustrate one of the promising optical design candidates in Figure\,\ref{fig:spectrograph}. Similar to the DESI's spectrograph, the fiber slit sits at the center of a dichroic. While this design increases the fabrication difficulty of the dichroic, it helps make the spectrograph more compact. It eliminates the need for the expensive and challenging Schmidt corrector that would be necessary for an off-axis fiber slit. The light from the fiber slit is reflected by a sperhical collimator mirror first. Unlike DESI, MUST's spectrograph will isolate the blue channel first after the first dichroic, and then split the beam into the red and NIR channels. Compared to DESI, this change will help improve the overall throughput of the blue channel, which is critical for the redshift measurements of $2.3<z<4.5$ high-$z$ galaxies for the Stage-V survey. Afterward, the beam for each channel is focused onto the CCD detector through a large FoV and fast focal ratio transmissive camera with an average throughput of $>$80\%. From blue to NIR, the average spectral resolutions of the three channels are around 2,500, 3,500, and 4,500, satisfying the design requirements.
 
    Note that, in this design, we avoid using CaF$_2$ aspheric lenses or doublets/triplets in the camera design to reduce manufacturing risk and cost. At the same time, this approach could potentially further improve spectral quality or resolution. Also, it is worth pointing out that the current design samples a spectrum using $\sim 4$ pixels. Compared to the optimal 3-pixel sampling, it results in a higher read-out noise contribution in the extracted spectrum. We are aware of possible solutions that could make the camera faster and reduce its FoV to achieve higher imaging quality. However, it often involves a drastic design change or introduces more lenses into the camera design, which decreases throughput, increases cost, and also makes camera alignment more difficult. The optical design of MUST's spectrograph has passed the conceptual design review, and we aim to explore a few more options before freezing the design.

    The 40 sets of spectrographs will be arranged in the spectrograph room $\sim 20$ meters below the telescope level in a circular layout. Due to the limited size of the room, we will place the spectrographs in stabilized racks with at least two levels. The spectrograph room will have overall temperature and humidity control to ensure the stability of the spectrographs during the operation.

\section{Concluding Remarks}
    \label{sec:conclusion}

    Astronomy has long stood at the frontier of human curiosity, driving our understanding of the Universe and our place within it. At its core, observational astronomy relies on the continual advancement of large-scale observing facilities and innovative technologies. In recent years, we have entered a transformative era marked by multi-messenger observations, the construction of ever-larger telescopes capable of grasping exquisite details of individual targets, and unprecedented wide-field imaging surveys that map the sky with remarkable depth and resolution. Yet, amid these achievements, a critical gap remains: the lack of highly multiplexed spectroscopic survey capabilities that can fully exploit the wealth of imaging data now available. Bridging this gap is not only a matter of technological advancement—it is essential for the future of cosmology, Galactic archaeology, and time-domain astrophysics, where statistical power and spectroscopic context are indispensable.

    In response to this need, we have proposed the development of a wide-field spectroscopic survey telescope, designated as MUST. This facility is designed to complement large general-purpose telescopes and imaging survey projects by extending the parameter space of observing capabilities and discovery potential. As the first Stage-V spectroscopic survey facility under construction, MUST has the potential to enable major scientific breakthroughs. Equipped to observe $\sim$20,000 targets simultaneously over a $\sim$5 deg$^2$ field using a 6.5-meter telescope, and to obtain low- to mid-resolution spectra ($R \sim 1900$--5000) spanning the entire optical wavelength range ($\lambda = 370$--960~nm), MUST will make transformative contributions to cosmology and fundamental physics. 

    In its preliminary design, MUST aims to conduct the first Stage-V spectroscopic cosmology survey in the early 2030s. Throughout a planned 8-year campaign, the facility will collect spectra and redshifts for more than 100 million galaxies and quasars across an area exceeding 10,000 deg$^2$. It will provide the most detailed 3D map of the $z<1.6$ Universe to date and obtain redshifts for 20-40 million objects in the $2<z<5$ range for the first time. These data will enable the most precise BAO measurements at $z<1.6$ and the strongest constraints on growth rate at $z>2$, offering crucial insights into the nature and evolution of dark energy and gravity. Combined with next-generation CMB data, the MUST survey will also deliver competitive constraints on primordial non-Gaussianity and neutrino mass.

    Beyond its primary cosmological goals, MUST's significantly improved spectroscopic survey efficiency (10$\times$ over Stage-IV), high target density ($>$4,000 deg$^{-2}$), and flexible, modular focal plane design position it as a powerful engine of discovery across a wide range of fields—including extragalactic astronomy, Galactic archaeology and near-field cosmology, stellar astrophysics, and time-domain science. MUST will also be an essential partner for ongoing and upcoming imaging surveys (e.g., CSST, Euclid, LSST), for following up transient events discovered by new time-domain facilities (e.g., Mozi, Si-Tian), and for identifying promising targets for detailed study with large-aperture telescopes.

    In summary, building on the legacy of previous spectroscopic surveys, MUST is poised to become a cornerstone of astronomical infrastructure: a flagship experiment in cosmology and fundamental physics, a flexible discovery machine across astrophysical disciplines, and a long-term platform for advancing technologies and instrumentation for future spectroscopic observations.

\Acknowledgements{
    The MUST project is supported by the Ministry of Science and Technology, China (Grant No. 2023YFA1605600) and the Ministry of Education, China. Yu Liu acknowledges the support from the National Science Foundation of China (No. 12303005) and the Shuimu Tsinghua Scholar Program (No. 2022SM173). Song Huang acknowledges the support from the National Natural Science Foundation of China (NSFC) Grant No. 12273015 \& No. 12433003 and the China Crewed Space Program through its Space Application System.
}

\InterestConflict{
     This paper was equally contributed to by Zheng Cai, Song Huang, Yu Liu, and Cheng Zhao. The authors declare that they have no conflict of interest.
}

\bibliography{main}
\bibliographystyle{scpma_short}

\end{multicols}
\end{document}